\documentclass[11pt]{article}
\usepackage[utf8]{inputenc}

\usepackage{fullpage}

\usepackage{amsmath}
\usepackage{amssymb}
\usepackage{amsthm}
\usepackage{authblk}

\usepackage{xfrac}
\usepackage{algpseudocode}
\usepackage{algorithm}
\usepackage{tikz}
\usetikzlibrary{positioning}
\usepackage{xcolor}

\usepackage{hyperref}
\usepackage{cleveref}
\usepackage{lipsum}

\title{A Critique of Du's ``A Polynomial-Time Algorithm for 3-\sat''\thanks{Supported in part by NSF grant 
		CCF-2006496.}}

\author{Yumeng He}
\author{Matan Kotler-Berkowitz}
\author{Harry Liuson}
\author{Zeyu~Nie}
\affil{Department of Computer Science\\University of Rochester\\Rochester, NY 14627, USA}

\newcommand{\p}{\ensuremath{{\rm P}}}
\newcommand{\np}{\ensuremath{{\rm NP}}}

\newcommand{\sat}{\ensuremath{{\rm SAT}}}

\newcommand{\true}{\textit{true}}

\newtheorem{definition}{Definition}

\date{April 5, 2024}

\begin{document}\sloppy

\maketitle

\begin{abstract}
In this paper, we examine the claims made by the paper ``A polynomial-time algorithm for 3-SAT''~\cite{du:t:poly-sat} by Lizhi Du. The paper claims to provide a polynomial-time algorithm for solving the NP-complete problem 3-$\sat$. In examining the paper's argument, we find a flaw in one of the main sections of its algorithm. We argue that this flaw causes the paper's algorithm to incorrectly decide that an infinite family of satisfiable 3-CNF boolean formulas are not satisfiable. Therefore, the paper does not establish that $\p = \np$.
\end{abstract}

\section{Introduction}
This critique provides an analysis of Lizhi Du's ``A polynomial-time algorithm for 3-$\sat$''~\cite{du:t:poly-sat}. The paper attempts to provide a polynomial-time algorithm that decides whether a 3-CNF boolean formula is satisfiable. This decision problem, known as 3-$\sat$, is a well-known NP-complete problem~\cite{karp:b:reducibilities}. Therefore, Du claims that their algorithm can be used to solve any problem in $\np$, in polynomial time. This claim has significant implications. Many widely-used algorithms have security claims that rely on the assumption that $\p \neq \np$, and many difficult problems may be easily solvable if $\p = \np$. For instance, if Du's claim is correct, most forms of cryptography may be breakable~\cite{mas-fab-marr:t:sat-cryto}, and problems such as protein folding and finding mathematical proofs may become solvable in polynomial time~\cite{berg-lei:t:protein-np-c}. We thus explore the algorithm that Du provides~\cite{du:t:poly-sat}. We show that the algorithm incorrectly decides that certain satisfiable 3-CNF boolean formulas are not satisfiable, which means that it is not a valid algorithm for 3-$\sat$.

In Section~\ref{s:analysis}, we present the relevant terminology from Du, and discuss how Du's algorithm purports to solve 3-$\sat$ in polynomial time. In particular, in Section~\ref{s:alg-1} we explain Du's ``Algorithm~1,'' which purports to identify so-called ``indirect contradiction pairs'': pairs of literals in a 3-CNF boolean formula for which no satisfying assignment exists that makes both literals true. Next, in Section~\ref{s:counterexample}, we provide a counterexample that demonstrates how Du's Algorithm 1 incorrectly decides that certain satisfiable 3-CNF boolean formulas are not satisfiable. Finally, we explain how this flaw means that Du's entire algorithm does not correctly decide 3-$\sat$, and thus Du's claim of $\p = \np$ does not follow from their arguments.

\section{Preliminaries}\label{s:pre}
In this paper, we assume basic familiarity with Turing machines and computation, including the complexity classes $\p$ and $\np$, and NP-complete problems. A boolean formula is in \textit{conjunctive normal form (CNF)} if it is the conjunction of one or more clauses, where each clause is the disjunction of zero or more \textit{literals}. Each literal can be either $x$ or $\bar x$ for any variable $x$. A boolean formula is in \textit{k-CNF form} if each clause of the formula has at most $k$ literals~\cite{karp:b:reducibilities}. For example, the following is a \textit{3-CNF} boolean formula:
\[(a \lor b \lor c) \land (d \lor \bar{b}).\]
The \textit{3-CNF satisfiability problem} is defined as determining whether a given 3-CNF boolean formula is satisfiable (i.e., has an assignment to the variables of the formula that makes the formula evaluate to true). The set 3-$\sat$ consists of all 3-CNF boolean formulas for which there exists a satisfying assignment.
3-$\sat$ is NP-complete~\cite{karp:b:reducibilities}. Thus the existence of a polynomial-time algorithm that decides 3-$\sat$ would imply $\p = \np$.

\section{Analysis}\label{s:analysis}
In this section, we introduce relevant terminology and algorithms from Du's paper\@. 

Given a 3-CNF boolean formula, Du's algorithm attempts to determine whether there is a satisfying assignment for that formula. Let $C_i = \{u_{i_1}, u_{i_2}, u_{i_3}\}$ refer to the $i$th clause of the 3-CNF formula, meaning that $C_i = u_{i_1}\lor u_{i_2}\lor u_{i_3}$. A ``unit'' is simply a literal in the clause. We note, before we introduce the following definitions, that Du's use of the term ``tree'' seems to refer to the way that the ``tree'' structure is constructed iteratively, with each ``layer'' being added one at a time. In this context, a layer corresponds to a clause in the original 3-$\sat$ instance, consisting of three or less ``units'' (literals). Du's ``trees'' have neither tree nor graph structure in the conventional sense.

Given a 3-CNF boolean formula, Du's algorithm transforms the formula into what the paper calls a ``standard checking tree'' through iteratively adding new clauses and deriving new ``contradiction pairs.''

\begin{definition}[Checking Tree]
    A checking tree is a sequence of layers. A checking tree with $k+1$ layers that are labeled $T_0,\ldots,T_k$, is denoted as $T^{(k)}$. Since checking trees are constructed iteratively, a checking tree may only have layers corresponding to some of the clauses in the original 3-SAT instance.
\end{definition}
\begin{definition}[Long Path]
For a checking tree $T$ with $m$ layers, a long path is a list of literals $P = (u_{1j_1}, u_{2j_2}, \ldots, u_{mj_m})$ such that $(\forall k\in\{1,2,\ldots,m\})[u_{kj_k} \in T_k\wedge j_k\in\{1,2,3\}]$ and no two literals in $P$ are negations of each other.
\end{definition}
\begin{definition}[Direct Contradiction Pair]\label{d:direct-contra}
A direct contradiction pair is a pair of literals $\{a, \bar{a}\}$.
\end{definition}
\begin{definition}[Indirect Contradiction Pair]\label{d:indirect-contra}
For a checking tree $T$, an indirect contradiction pair is a pair of literals $\{a,b\}$ such that there does not exist a long path $P$ of $T$ such that both $a,b \in P$.
\end{definition}

\begin{definition}[Standard Checking Tree]\label{d:std-checking-tree}
A standard checking tree is a pair $(T,C)$ where $T$ is a checking tree and $C$ is the set of contradiction pairs of $T$. Contradiction pairs may be either direct or indirect (Definitions~\ref{d:direct-contra} and~\ref{d:indirect-contra}).
\end{definition}

In essence, a long path is a satisfying assignment of the partial 3-$\sat$ instance represented by $T$, with its clauses corresponding to the layers of $T$. The goal of Du's algorithm is to find a long path through the standard checking tree that corresponds to the original 3-CNF formula.


The general process that Du gives to solve 3-$\sat$ is to iteratively construct the standard checking tree by adding one layer at a time. After the algorithm calculates all the contradiction pairs of $T^{(k)}$ for some natural number $k \geq 1$, it adds a new layer corresponding to $C_{k+1}$ to form $T^{(k+1)}$, and calculates the additional contradiction pairs of $T^{(k+1)}$. The algorithm does this by using a destroyed checking tree, which we define below.

\subsection{Destroyed Checking Tree}\label{s:dct}

\begin{definition}[Destroyed Checking Tree]
    A destroyed checking tree $D(T, S)$ of a standard checking tree $T$ and some subset of literals $S$ is $T$ such that for all literals $s \in S$, if $s = x$ for some variable $x$, then, then all the occurences of $\bar{x}$ are removed from $T$, and if $s = \bar{x}$ for some variable $x$, then all occurences of $x$ are removed from $T$. This corresponds to a modified problem with the restriction that the literals in $S$ cannot be set to true.
\end{definition}
Let $x,y$ be any pair of literals in the layers of $T^{(k)}$. To determine whether $(x,y)$ is a new indirect contradiction pair, for all $v \in C_{k+1}$, Du's algorithm constructs a destroyed checking tree $D(T^{(k)}, \{x, y, v\})$. If for all such $v \in C_{k+1}$, the destroyed checking tree is unsatisfiable, then $(x,y)$ must be a contradiction pair in $T^{(k+1)}$, as a long path through $T^{(k+1)}$ must have a literal in the last layer, $C_{k+1}$.

This requires that we determine the satisfiability of the destroyed checking tree. Du's algorithm first ``repairs" the destroyed checking tree to a standard checking tree using Algorithm 1. When a destroyed checking tree is created, it is comprised of the set of contradiction pairs from the standard checking tree from which it was created. However, these contradiction pairs may no longer be valid in the destroyed checking tree, as the destroyed literals impose additional constraints on the possible solutions to the problem. As such, Du's paper uses Algorithm 1 to augment the set of contradiction pairs and ``repair" the checking tree to obtain an (ostensibly) updated, correct set of contradiction pairs for the destroyed checking tree.

It then uses a somewhat convoluted procedure, Algorithm 2, which we will not discuss in depth here, in order to determine satisfiability. We observe that this procedure can actually be simplified, as given a standard checking tree $T$ and a set of contradiction pairs $R$, $T$ is unsatisfiable if and only if $R$ contains every pair of units in (some layers of) $T$. Then we can check the satisfiability of the destroyed checking tree by invoking Algorithm 1. 

\subsection{Algorithm 1}\label{s:alg-1}
Algorithm 1 takes a destroyed checking tree $D(T, S)$ as input. Let $\ell_2$ be the set of clauses with one or two literals, created as a result of deleting literals. Let $\ell_3$ be the set of clauses with three literals.

For each unit $x$ that is a member of a clause in $\ell_3$, the algorithm calculates its \emph{useful units} $U_x$ as the set of literals in clauses in $\ell_2$ that are in a satisfying partial assignment of $\ell_2$ consistent with $x$. This is decidable in polynomial time, since 2-$\sat$ $\in$ P.

The issue arises in Step 3: ``for each unit $u$’s each useful unit (sic), if in another 3-unit layer, the units that do not destroy $u$ do not have this useful unit, $u$ also has to lose this useful unit''~\cite{du:t:poly-sat}.

``Destroy'' here is not entirely clearly-defined. We interpret this as:
For all $(x,y)$ in distinct 3-unit layers that are not contradiction pairs of $T$, the useful units of $x$ and $y$ are set as the intersection between the useful units of $x$ and $y$. Section~\ref{s:counterexample} demonstrates how Step 3 causes the algorithm to incorrectly decide that certain 3-CNF formulas are unsatisfiable.

\section{A Counterexample}\label{s:counterexample}
Consider an instance of a standard checking tree, where $C_1 \dots C_n$ are some arbitrary clauses satisfying our assumptions below.
\begin{multline}\label{f:destroyed-ct}
\\
    s \lor t \lor \bar{c} \\
    C_1\\
    \vdots\\
    C_n\\
    \bar{s} \lor \bar{t} \lor r.\\ 
\end{multline}
Now, suppose we are trying to add the clause $a \lor b \lor c$ to the tree.

Assume the existence of some unit $\alpha$ such that $(a, \alpha)$ and $(b, \alpha)$ are contradiction pairs. Additionally suppose that all long paths contain $c, \alpha$, and that $(s, \bar{t})$ and $(t, \bar{s})$ are not contradiction pairs, and suppose that there exists at least one long path.

Then to check if $c, \alpha$ are a contradiction pair, we delete $\bar{c}$ and $\bar{\alpha}$ from the tree. The resulting destroyed checking tree has the form:

\begin{multline}\label{f:destroyed-ct}
\\
    s \lor t \\
    C'_1\\
    \vdots\\
    C'_n\\
    \bar{s} \lor \bar{t} \lor r.\\ 
\end{multline}

If $(s, \bar{t})$ and $(t, \bar{s})$ are not contradiction pairs, and $\alpha$ is in all long paths, then there must be some long path which has both $\alpha$ and $\bar{s}$ and another one which has both $\alpha$ and $\bar{t}$. Therefore $(\alpha, \bar{s})$ and $(\alpha, \bar{t})$ are not indirect contradiction pairs. That is, they do not destroy each other. Clearly, $t$ is $\bar{s}$'s only useful unit and $s$ is $\bar{t}$'s only useful unit. Then Step 3 of Algorithm 1 causes $s$ and $t$ to be removed from $\alpha$'s useful units. This causes $\alpha$ to be deleted in Step 7 since it has a layer with zero useful units.

By our assumption that all valid long paths must contain $\alpha$, the algorithm then concludes that there are no solutions and that $(c, \alpha)$ must be a contradiction pair.

This demonstrates one failure case of Algorithm 1. It can incorrectly conclude that a destroyed checking tree has no solutions, leading to pairs incorrectly being designated as indirect contradiction pairs. It is clear that incorrectly labeling contradiction pairs may lead to the algorithm failing to output the correct answer. For example, in the above case, if there was a single unique solution requiring $c$ and $\alpha$ to be assigned $\true$, then the algorithm would fail.

Additionally, the clauses $C_1, \ldots, C_2$ may be composed of any arbitrary literals that satisfy our assumptions. Therefore, there is an infinite family of 3-$\sat$ instances that Du's algorithm decides incorrectly.







\section{Conclusion}\label{s:conclusion}
Du presents a complex algorithm that claims to decide in polynomial time whether a 3-$\sat$ instance is satisfiable. Du's algorithm is presented in multiple parts, with each part relying on the output of the previous part. The only part of Du's paper that does not rely on another algorithm is Algorithm 1. If we assume that Algorithm 1 works correctly, the rest of Du's algorithm may indeed be able to decide 3-$\sat$ in polynomial time. However, as demonstrated above, Algorithm 1 does not work correctly: There are satisfiable instances of 3-$\sat$ that Algorithm 1 will identify as being unsatisfiable. Since the entirety of Du's algorithm relies on Algorithm 1, this flaw in Algorithm 1 is fatal to the rest of Du's algorithm. Thus Du's paper has not provided an algorithm that correctly decides an NP-complete problem in polynomial time. As a result, the claim that $\p = \np$ does not follow from the paper's argument.

\paragraph{Acknowledgments}

We would like to thank
Michael C. Chavrimootoo,
Lane A. Hemaspaandra,
Michael P. Reidy, and
Eliot J. Smith
for their helpful comments on prior drafts.
The authors are responsible for any remaining errors.

\bibliographystyle{alpha}
\bibliography{gry-reu,local_refs}

\end{document}